\documentclass[twocolumn, preprintnumbers, 
endnote,nofootinbib,prl,9pt]{revtex4}

\usepackage{epsfig,subfigure}

\usepackage{epstopdf}

\usepackage[utf8]{inputenc}
\usepackage{graphicx}
\usepackage{amssymb}
\usepackage{amsxtra}
\usepackage{amsmath}
\usepackage{booktabs,multirow,tabularx}
\usepackage{slashed}
\usepackage{float}
\usepackage{placeins}
\usepackage{rotating}
\usepackage{lscape}
\usepackage{color}
\usepackage{hyperref}

\usepackage[Q=yes,pverb-linebreak=no]{examplep}

\DeclareGraphicsExtensions{.eps}
\graphicspath{{./}}

\newcommand{\vev}[1]{\langle {#1} \rangle}
\newcommand{\lsim}{\lesssim}
\newcommand{\gsim}{\gtrsim}

\newcommand{\nn}{\nonumber}

\newcommand{\gev}{\,\textrm{GeV}}

\newcommand{\tev}{\,\textrm{TeV}}

\def\beq{\begin{equation}}
\def\bea{\begin{eqnarray}}
\def\eeq{\end{equation}}
\def\eea{\end{eqnarray}}
\def\beqnl{\begin{align}}
\def\endal{\end{align}}

\newcommand{\mul}{\mathcal{M}_\Lambda}
\newcommand{\sraise}{\uparrow}
\newcommand{\slower}{\downarrow}
\newcommand{\sraiselower}{\updownarrow}
\newcommand{\spup}{\land}
\newcommand{\sdown}{\lor}

\DeclareFontFamily{U}{cbgreek}{}
\DeclareFontShape{U}{cbgreek}{m}{n}{
        <-6>    grmn0500
        <6-7>   grmn0600
        <7-8>   grmn0700
        <8-9>   grmn0800
        <9-10>  grmn0900
        <10-12> grmn1000
        <12-17> grmn1200
        <17->   grmn1728
      }{}
\DeclareFontShape{U}{cbgreek}{bx}{n}{
        <-6>    grxn0500
        <6-7>   grxn0600
        <7-8>   grxn0700
        <8-9>   grxn0800
        <9-10>  grxn0900
        <10-12> grxn1000
        <12-17> grxn1200
        <17->   grxn1728
      }{}

\DeclareRobustCommand{\Qoppa}{%
  \text{\usefont{U}{cbgreek}{\normalorbold}{n}\symbol{21}}%
}
\makeatletter
\newcommand{\normalorbold}{%
  \ifnum\pdf@strcmp{\math@version}{bold}=\z@ bx\else m\fi
}
\makeatother

\begin{document}

\title{Exo-Higgs at 750 GeV and Genesis of Baryons}

\author{Hooman Davoudiasl\footnote{email: hooman@bnl.gov}
}

\author{Pier Paolo Giardino\footnote{email: pgiardino@bnl.gov}
}

\author{Cen Zhang\footnote{email: cenzhang@bnl.gov}
}

\affiliation{Department of Physics, Brookhaven National Laboratory,
Upton, NY 11973, USA}

\begin{abstract}

We propose that the diphoton excess at 750~GeV reported by ATLAS and CMS
is due to the decay of an {\it exo-Higgs} scalar $\eta$
associated with the breaking of a new $SU(2)_e$ symmetry, dubbed {\it exo-spin}.
New fermions, {\it exo-quarks} and {\it exo-leptons}, get TeV-scale masses
through Yukawa couplings with $\eta$ and generate its couplings to gluons and
photons at 1-loop.  The matter content of our model yields a $B-L$ anomaly
under $SU(2)_e$, whose breaking we assume entails a first order phase transition.  A
non-trivial $B-L$ asymmetry may therefore be generated in the early universe, potentially providing
a baryogenesis mechanism through the Standard Model (SM) sphaleron processes.  The spontaneous
breaking of $SU(2)_e$ can in principle directly lead to electroweak symmetry breaking, thereby
accounting for the proximity of the mass scales of the SM Higgs and the exo-Higgs. Our model can be distinguished from those comprising a singlet scalar and
vector fermions by the discovery of TeV scale exo-vector bosons,
corresponding to the broken $SU(2)_e$ generators, at the LHC.

\end{abstract}

\maketitle

\section{Introduction\label{sec:intro}}

The diphoton excess at the LHC, reported by both the ATLAS and CMS \cite{ATLAS,CMS:2016owr} 
collaborations at about $750\gev$, has been the subject of
a large number of papers over the past several months\footnote{A large number of papers have been written on this subject since the initial announcement of the excess, as can be seen from 
the citations of Refs.~\cite{ATLAS,CMS:2016owr} \label{fn:one}}.  While the significance of the excess is not at the
discovery level yet, its appearance in both experiments, persistence upon further analysis, and the nature of the final state
provide some ground for cautious optimism that it may be a real signal of new physics.  One is then compelled to ask
what the underlying new physics can be.

Many ideas have been entertained and cover a multitude of possibilities\footnote{See footnote (\ref{fn:one})}. However,
among them, the possibility of a scalar resonance with a mass of $750\gev$, produced via gluon fusion and
decaying into photons, both at 1-loop level, represents one of the most straightforward scenarios (See, for example, Refs.~\cite{Knapen:2015dap,Franceschini:2015kwy,McDermott:2015sck,Falkowski:2015swt,Aloni:2015mxa}).  The gluon
initial states are well-motivated, as their corresponding luminosity gets enhanced
much more than that for the quarks
with center of mass energy of collisions, greatly reducing tension with the LHC Run 1 data.  Here, the particles
that mediate the loop-generated couplings of the scalar are generally assumed to be heavy vector-like fermions that
carry color and charge, as mediation by lighter states, such as those in the Standard Model (SM), would provide tree-level decay modes that would
make the requisite diphoton signal strength hard to explain.

The above simple setup would then suffice to account for the key features of the excess, as they
are currently known.  However, one may then inquire how the new states
may fit within a larger picture of particle physics.  Obviously, this question could be answered in a variety
of ways, depending on one's view of fundamental physics and its open problems.

In this work, we entertain the possibility that the $750\gev$ resonance is a scalar remnant of a TeV-scale
Higgs mechanism responsible for the spontaneous breaking of a new $SU(2)_e$ gauge symmetry that we refer to as {\it exo-spin}
({\it exo: outside}, in Greek).  None of the SM fields carry $SU(2)_e$, however there are new fermions charged under this symmetry, as well as
under the SM $SU(3)_c$ color and hypercharge $U(1)_Y$.  We will refer to the new color charged fermions
as {\it exo-quarks}, while those that only carry hypercharge are referred to as {\it exo-leptons}.
These fermions get their masses through Yukawa coupling to a doublet {\it exo-Higgs}
whose vacuum expectation value (vev) breaks the $SU(2)_e$ symmetry.

Our proposed setup is motivated by the natural assumption that a particle whose properties
are reminiscent of the SM Higgs is perhaps best thought of as a Higgs boson that breaks a new symmetry (For a sample of works that consider a Higgs field interpretation of the excess, see Refs.~\cite{Angelescu:2015uiz,Gupta:2015zzs,Chang:2015bzc,Kaneta:2015qpf,Duerr:2016eme}).
The simplicity and minimal nature of the $SU(2)_e$ group make it a compelling choice, however we
go a step further and will assume that it is responsible for generating the non-zero baryon asymmetry
in the early universe, thereby addressing one of the main open questions in cosmology and particle physics. More specifically,
we choose our exo-fermion quantum numbers such that $B-L$, with $B$ baryon number and $L$ lepton number, is {\it anomalous}
under $SU(2)_e$.  One can then envision that if $SU(2)_e$ breaking in the early universe entailed a
{\it first order phase transition}, at a temperature $T\sim 1$~TeV, the associated departure from equilibrium could lead to
the appearance of non-zero $B-L$ that would then get processed into the cosmic baryon asymmetry by the
SM sphalerons \cite{Harvey:1990qw} (see also Refs.~\cite{Shu:2006mm,Agashe:2010gt,Walker:2012ka}).  This is similar to the scenarios envisioned for electroweak baryogenesis that would
require a strongly first order electroweak phase transition, which is however not realized by the SM Higgs.  We then
require that the coupling of the $SU(2)_e$  be large enough that, while perturbative, would still lead to a first order
phase transition.

In the heavy exo-fermion limit, the diphoton signal strength is largely a function of
the vev of the exo-Higgs doublet, which can then be fixed.  Hence, the exo-Higgs potential parameters can be obtained
for a given signal strength.  We will also assume that the the SM Higgs portal coupling with
exo-Higgs generates the SM Higgs mass parameter after $SU(2)_e$ breaking. This not only reduces the
number of input parameters in the model, but also provides an explanation of the relative proximity of the
the Higgs and exo-Higgs masses.  
We will next introduce the ingredients of our model and discuss
its potential relevance to baryogenesis.

\section{The Model}
In this section we will describe the main features of our model.
We assume the existence of a new gauge symmetry $SU(2)_e$, completely broken by a Higgs field $\eta$, under which the SM fermions are singlets.
As per the usual Higgs mechanism, three degrees of freedom of 
$\eta$ give masses to the three gauge bosons $\omega_1$,  $\omega_2$ and $\omega_3$ 
associated with $SU(2)_e$, while we identify the fourth with the $750\gev$ resonance. Unlike in the SM,  $\omega_{1,2,3}$ are degenerate in mass. 
We also introduce new fermions, charged under $SU(2)_e$ and the SM gauge group, that acquire mass through Yukawa couplings with the exo-Higgs $\eta$.
Following the SM naming rule we call the fermions in a triplet of $SU(3)_c$ exo-quarks 
$\Qoppa$ (archaic Greek letter pronounced {\it Koppa}), and the fermions that are singlets of $SU(3)_c$ exo-leptons $\Lambda$. The choice of possible quantum numbers is limited by requiring that the theory is free of 
gauge anomalies. Since the exo-fermions are vector-like under the SM gauge group, freedom from anomalies is trivially satisfied for that sector and the only non-trivial anomalies are the Adler-Bell-Jackiw with one $U(1)_Y$ and two $SU(2)_e$ bosons and the Witten anomaly \cite{Witten:1982fp}.  
We found the following anomaly free choice of quantum numbers under $SU(2)_e \otimes SU(3)_c \otimes SU(2)_L \otimes U(1)_Y$ particularly interesting:
\bea
\label{eq:qn}
\Qoppa_L&=&(2,3,1,-\frac{1}{3})\\
\Qoppa_R&=&(1,3,1,-\frac{1}{3})    \quad (\times 2) \nonumber\\
\Lambda_L&=&(1,1,1,-1)  \quad (\times 2)   \nonumber\\
\Lambda_R&=&(2,1,1,-1)\nonumber,
\eea
where for the upper (lower) component of a $\Qoppa_L$ doublet we have a corresponding $\Qoppa_R^{\spup}$ ($\Qoppa_R^{\sdown}$). We adopt a similar notation for the $\Lambda_R$ doublet. Note that, since the exo-fermions are always singlets under $SU(2)_L$, the $U(1)_Y$ charge coincide with the electric charge.
Lastly, we consider three generations of exo-fermions, $\Qoppa^{\{1,2,3\}}$ and $\Lambda^{\{1,2,3\}}$. This completes our definitions for the field content of our model.

The Lagrangian is the sum of three contributions:
\bea
\mathcal{L}=\mathcal{L}_{SM}+\mathcal{L}_e+\mathcal{L}_m,
\label{eq:cL}
\eea
where the first term is the SM Lagrangian without the Higgs doublet mass term $\mu_H^2 H^\dagger H$, the second is the Lagrangian of the exo-sector, and in the third we have the terms of mixing between SM and exo-sector.
The second term of Eq.~(\ref{eq:cL}) is
\bea
\label{eq:Le}
\mathcal{L}_e&=&-\frac{1}{4}\omega^a_{\mu\nu}\omega^{\mu\nu}_a+ (D_\mu \eta)^\dagger(D^\mu\eta)+\mu_\eta^2 \eta^\dagger\eta-\lambda_\eta |\eta^\dagger\eta|^2\nonumber\\
&+&i\bar{\Qoppa}_L\slashed{D} \Qoppa_L+i\bar{\Qoppa}_R\slashed{D} \Qoppa_R+i\bar{\Lambda}_L\slashed{D} \Lambda_L+i\bar{\Lambda}_R\slashed{D}\Lambda_R\nonumber\\
&-&Y_\Qoppa^{\sdown;i,j} \eta \bar{\Qoppa}_L^i \Qoppa_R^{\sdown;j}-Y_\Qoppa^{\spup;i,j} \tilde{\eta} \bar{\Qoppa}_L^i \Qoppa_R^{\spup;j}\nonumber\\
&-&Y_\Lambda^{\sdown;i,j} \eta \bar{\Lambda}_R^i \Lambda_L^{\sdown;j}-Y_\Lambda^{\spup;i,j} \tilde{\eta} \bar{\Lambda}_R^i \Lambda_L^{\spup;j},
\eea
plus the usual gauge fixing and Fadeev-Popov ghost terms. The indices $i,j$ refer to different generations. As in the SM we can rotate the fields to a mass basis, and generate the exo-sector counter-parts of the CKM and PMNS matrices.

We now discuss the mixing terms. The mixing Lagrangian can be written in general as
\bea
\label{eq:Lm}
\mathcal{L}_m&=&  2 k_{\eta H} \eta^\dagger\eta H^\dagger H\\
&-&Y_{\Qoppa q}^{\sdown;i,j} \eta\, \bar{\Qoppa}_L^i d_R^j- Y_{\Qoppa q}^{\spup;i,j}\  \tilde{\eta}\, \bar{\Qoppa}_L^i d_R^j\nn\\
&-&Y_{q\Qoppa}^{\sdown;i,j} H \bar{q}_L^i  \Qoppa_R^{\sdown;j}-Y_{q\Qoppa}^{\spup;i,j} H \bar{q}_L^i  \Qoppa_R^{\spup;j}\nn\\
&-&\mul^{\spup;i,j} \bar{\Lambda}_{L}^{\spup;i} e_R^j-\mul^{\sdown;i,j} \bar{\Lambda}_{L}^{\sdown;i} e_R^j,\nn
\eea
where $q_L$ is the left-handed quark doublet, $d_R$ is a right-handed down-type quark and $e_R$ is a right-handed charged lepton.
The first term of eq. (\ref{eq:Lm}) is the mixing between the Higgs fields of
the two sectors (we will be interested in values of $k_{\eta H}^2 \ll \lambda_H \lambda_\eta$ and 
hence the negative sign of this interaction does not yield an unstable potential). We fix the value of $k_{\eta H}$ imposing that $k_{\eta H}
v_\eta^2 = \mu_H^2 $ where $v_\eta$ is the vev of the
$\eta$ field, $\vev{\eta}=v_\eta/\sqrt{2}$, and $\mu_H=\sqrt{\lambda v_H^2}$ where
$v_H$ is the vev of the SM Higgs doublet. In this way, we can justify the
proximity between the breaking scale of $SU(2)_e$ and the vev of SM Higgs doublet. This is
not a requirement of our model, but the predicted value of $k_{\eta H}$ sits
well within the phenomenological constraints, as we will see later.

In the second line of Eq.~(\ref{eq:qn}), $\Qoppa_R$ has the same quantum numbers as a $d_R$, so it can couple to $q_L$ with a Yukawa interaction mediated by the SM Higgs. In the same way $d_R$ can couple to $\Qoppa_L$ as shown in the second and third lines of Eq.~(\ref{eq:Lm}). The Yukawa matrices will generically have off-diagonal terms, that can produce Flavor Changing Neutral Currents (FCNC). However, we can set the off-diagonal terms as small as we want without altering the main purpose of this work, and, for simplicity, we will consider all the Yukawa matrices in Eq.~(\ref{eq:Lm}) to be diagonal.

The lepton sector is peculiar since the quantum numbers of $\Lambda_L$ allow us to write a mixing term where $\mul$ is a mass parameter not directly related to the other scales of the theory ($v_H$ and $v_\eta$). As we will see in the phenomenology section, its value should be at least one order of magnitude smaller than $v_H$. However, that relatively 
small value can be justified if this interaction descends from a higher energy theory 
where heavier degrees of freedom have been integrated out.
As before, off-diagonal $\mul$ terms can generate FCNC in the lepton sector, but we can assume $\mul$ to be diagonal for 
the purposes of this paper.

The off-diagonal terms in the exo-quark Yukawa matrices in Eq. (\ref{eq:Le}) can also be a source of FCNCs. However the mixing between SM-quarks and exo-quarks can be set small to avoid strong bounds on the CKM of the exo-sector. In any case, for simplicity, we will consider a limit where
all the off-diagonal terms are zero, the elements of the same exo-doublet are degenerate in mass and two of the exo-quark generations have the same mass. It has to be noted, however, that in general a more complex Yukawa sector for the exo-quarks is desirable, since it can be a source of the extra $CP$-violation needed for baryogenesis.  A similar reasoning applies to the exo-lepton sector, where we will set the Yukawa matrices diagonal and all the exo-leptons degenerate in the mass.

\begin{figure}[t]
\begin{center}
\includegraphics[width=0.45\textwidth]{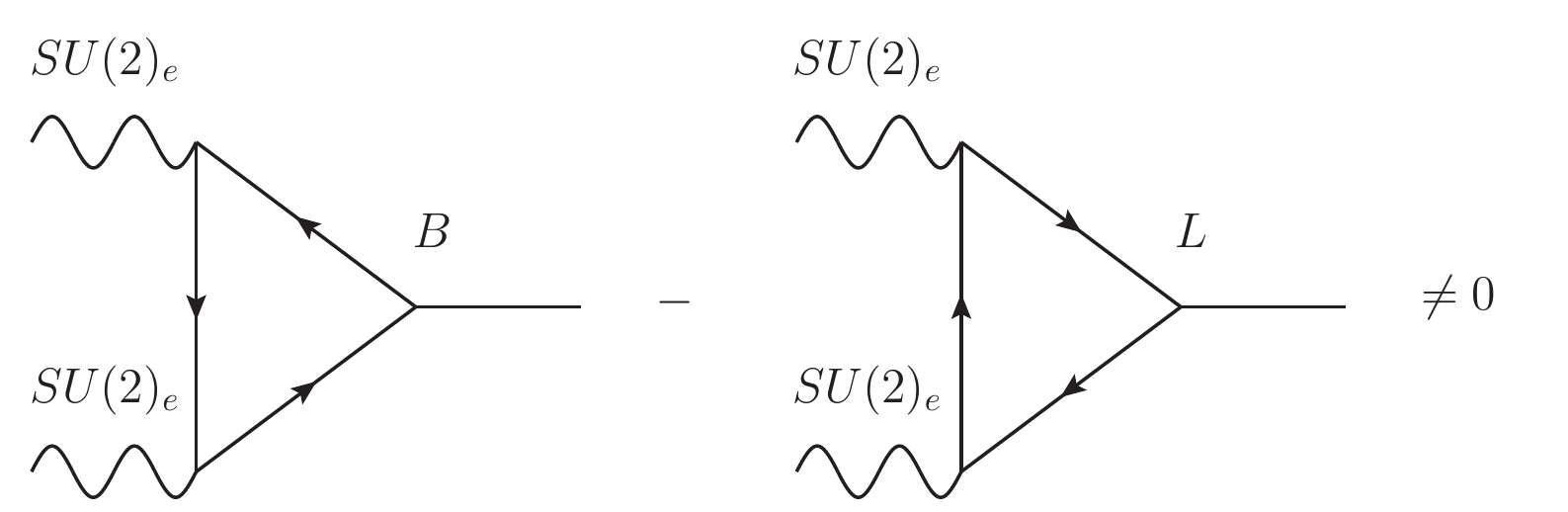}
\end{center}
\caption{Triangular anomaly for $B-L$. Although the field content of the
exo-sector is similar to the one of the SM, the exo-leptons are right-handed
doublets, so their contribution has a sign opposite to that of the exo-quarks.}
\label{fig:BL}
\end{figure}

It is interesting to note that, at tree level, the Lagrangian in Eq.~(\ref{eq:cL}) preserves $B$ and $L$, once we assign to the exo-quarks and the exo-leptons the quantum numbers $B=\frac{1}{3}$ and $L=1$, respectively. However once we compute the triangular anomaly in Fig. \ref{fig:BL}, the resulting $B-L$ current is anomalous, while the $B+L$ is preserved.

\section{Connection to Cosmology}

As discussed in the last section, $B-L$ number is anomalous under our
$SU(2)_e$. In what follows, we will argue that this anomaly offers a
possibility to address an important open question in cosmology, that is the 
origin of baryon asymmetry in the universe.  To address this question, one
needs to introduce a baryogenesis mechanism that satisfies Sakharov's
criteria \cite{Sakharov:1967dj}: {\it (i)} baryon number violation, {\it (ii)} $C$ and $CP$
violation, and {\it (iii)} departure from equilibrium.  In the SM, {\it
(i)} is provided by sphaleron processes at temperatures $T \gtrsim
100$~GeV before spontaneous electroweak symmetry breaking.  Both $C$ and
$CP$ violation are present in the SM, but the amount of $CP$ violation is
too small.  Condition {\it (iii)} would have required a first order
electroweak phase transition, which is not feasible with the SM Higgs
potential.

Various extensions of the SM have been proposed in order to supplement its
shortcomings in the context of electroweak baryogenesis.  In particular,
one can entertain the possibility that an initial $B-L$ number, which is
respected by all SM interactions, was present well before electroweak
symmetry breaking took place.  The SM sphalerons would then process the
$B-L$ asymmetry into $\Delta B\neq 0$ and $\Delta L\neq 0$ asymmetries
{\it in equilibrium}.  A well-motivated scenario of this kind, referred to
as {\it leptogenesis} \cite{Fukugita:1986hr}, employs heavy Majorana neutrinos that are required to
implement a seesaw mechanism for generating light SM neutrino masses.
While an interesting idea, leptogenesis typically requires that Majorana states
appear at scales $\gg 1$~TeV, well beyond the reach of direct discovery.
Hence, this idea is only indirectly testable.

Here, we propose that the $B-L$ anomaly in our model can lead to the
generation of $\Delta(B-L)\neq 0$ if $SU(2)_e$ breaking at $T\sim 1$~TeV
involves a first order phase transition.  This is in analogy with
electroweak baryogenesis mechanisms, where a strong transition would have
generated departure from equilibrium as required for a baryon asymmetry.
The generation of a $B-L$ asymmetry would also require sources of $CP$
violation, which our model would readily provide once generally complex Yukawa couplings are assumed.  
This motivates us to consider model parameters
that support a first order $SU(2)_e$ phase transition.  The key reason the
SM cannot afford this possibility is that the finite temperature effective potential
for the Higgs has a thermally generated {\it cubic} term that is too
small.  This term can lead to the appearance of a requisite barrier in the
potential during the phase transition.

By analogy with the SM case, we see that the coefficient of the relevant
cubic term for the $SU(2)_e$ transition is (see, for example,
Ref.~\cite{Quiros:1999jp})
\beq
E= \frac{3 \, m_\omega^3}{4\pi v_\eta^3} =
\frac{3 \, g_e^3}{32\pi}\,.
\label{cubic}
\eeq
A strong first order phase transition is one whose
order parameter at the critical temperature of the transition $T_c$
satisfies $\eta(T_c)/T_c \gsim 1$, where $\eta(T_c)$ is the exo-Higgs
field value at the local minimum of the effective potential for $T=T_c$.
One can show \cite{Quiros:1999jp} that this requirement then implies $2
E/\lambda_\eta(T_c) \gsim 1$, where $\lambda_\eta(T_c)$ is the quartic
self-coupling of the $\eta$ at $T_c$.

Using the approximation $\lambda_\eta(T_c) \approx \lambda_\eta$, the
condition for a strong
first order phase transition in our model can then be written as
\beq
\frac{3 \, g_e^3}{16 \pi\, \lambda_\eta} \gsim 1\,.
\label{fopt-cond}
\eeq
Later, we will show that the signal strength suggested by the diphoton
excess is, given our choice of model parameters and ingredients,
mainly sensitive
to $\vev{\eta}$.  Hence, for a given signal strength, and assuming that
the
exo-Higgs mass is 750~GeV, one can determine $\lambda_\eta$ within our
setup.  We can
then use Eq.~(\ref{fopt-cond}) to derive a lower bound on $g_e$, motivated
by the
possibility of explaining the baryon asymmetry of the universe, as
explained above.  Such
a baryogenesis mechanism will have the great advantage of being testable
at the LHC and
future high energy colliders, given that it is based on physics at or near
the TeV scale.

\section{Signal strength and parameters}

In this section, we will discuss the signal strength and its implications for our
parameter space.  All the cross sections, decay rates, and branching ratios,
appearing in this and the next sections, are obtained using {\sc
MadGraph5\_aMC@NLO} \cite{Alwall:2014hca} and {\sc MadWidth}
\cite{Alwall:2014bza} with a UFO model
\cite{Degrande:2011ua} made with the {\sc FeynRules} package
\cite{Alloul:2013bka}.  Loop-induced processes are computed following
Ref.~\cite{Hirschi:2015iia}, and the corresponding counter term is computed with
{\sc NLOCT} \cite{Degrande:2014vpa}.  In all simulations we use NNPDF2.3 parton
distributions \cite{Ball:2013hta}, except for the 750 GeV signal, where we use
CT14nlo \cite{Dulat:2015mca} to be consistent with Ref.~\cite{Liu:2016mpd}, so
that we can apply the NLO+NNLL $K$ factor computed with the same setup.

The signal strength is mostly sensitive to the vev of the exo-Higgs doublet.
Similarly to the SM, in the heavy fermion mass limit, the loop-induced
$gg\to\eta$ and $\eta\to\gamma\gamma$ are described by the following
dimension-five operators:
\begin{flalign}
	\mathcal{L}=\frac{\alpha_s}{3\pi v_\eta}N_{\Qoppa}
	\frac{1}{4}G_{\mu\nu}^AG^{A\mu\nu}\eta
	+\frac{2\alpha}{3\pi v_\eta}\sum_{\Qoppa,\Lambda}N_cQ_f^2
	\frac{1}{4}F_{\mu\nu}F^{\mu\nu}\eta
\end{flalign}
With our setup, $v_\eta\sim1$ TeV will roughly produce the observed signal
strength.  Other parameters, like mixing angles and exo-Yukawas, could
also affect the signal strength, by modifying the decay modes of $\eta$ and
the fermion masses running in the loop.

To facilitate a more concrete discussion, let us consider the following benchmark
point  
\begin{flalign}
\begin{array}{lcll}
v_\eta & = & 1.2 \tev & \text{Vev of the $\eta$ field}\\
m_\Qoppa^> & = & 800 \gev & \text{Mass of the heavier $\Qoppa$}\\
m_\Qoppa^< & = & 500 \gev & \text{Mass of the lighter $\Qoppa$'s}\\
m_\Lambda & = & 380 \gev & \text{Mass of $\Lambda$'s}\\
\mul & = & 1 \gev & \text{$\Lambda-l$ mixing mass parameter}\\
 \theta^\Qoppa_{L,R}  & = & 10^{-3}  & \text{$\Qoppa-q$ mixing angles}\\
 g_e & = & 2 & \text{$SU(2)_e$ gauge coupling}\\
\end{array}
\label{eq:benchmark}
\end{flalign}
With these parameters, we find that the cross section for the 750 GeV signal is
about 4.1 fb, taking into account an NLO+NNLL $K$ factor of 1.56 from
Ref.~\cite{Liu:2016mpd}.  This value should be compared with the weighted
average deduced from the ATLAS \cite{ATLAS} and CMS \cite{CMS:2016owr} data $4.6\pm 1.3$ fb.

\begin{table}[h]
	\begin{tabular}{cc}
		Mode & BR
		\\\hline
		$gg$ & 82.1\%
		\\
		$W^+W^-$ & 7.4\%
		\\
		$HH$ & 4.2\%
		\\
		$ZZ$ & 3.7\%
		\\
		$t\bar t$ & 1.8\%
		\\
		$\gamma\gamma$ & 0.81\%
		\\
		$\gamma Z$ & 0.47\%
		\\
		$l\Lambda$ & 0.12\%
		\\\hline\hline
		$\Gamma_\eta$ & 0.060 GeV
	\end{tabular}
	\caption{Main branching ratios and the total width $\Gamma_\eta$ of the exo-Higgs $\eta$.  Channels with
		lower than 0.01\% branching ratio are not displayed.  The
		values in the table correspond to the benchmark point in
		Eq.~(\ref{eq:benchmark}).\label{tab:1}}
\end{table}

The $2k_{\eta H}\eta^\dagger\eta H^\dagger H$ term induces a mixing between
$H$ and $\eta$. With our assumption, namely that $v_\eta$ provides the $\mu_H$
term of the SM and leads to electroweak symmetry breaking, this mixing is about
$\sin\theta_{\eta H}=0.006$.  As a consequence, $\eta$ could decay to SM particles, such as
$W^+W^-$, $ZZ$, $t\bar t$, and $HH$, through the Higgs portal, and these channels
can be searched for in the future.  They also affect the value of
$v_\eta$, by diluting our signal strength by about $\sim15\%$. With mixing
higher than this value it will be difficult to achieve the desired signal strength,
so in this sense $\theta_{\eta H}$ is bounded from above.

The main branching fractions of $\eta$ and its total width $\Gamma_\eta$ are given in Table \ref{tab:1}. We find that  the implied signal strengths for these channels are not in conflict with existing constraints from the LHC Run 1 data.  As can be seen from the 
table, the branching fraction for $\eta \to ZZ$ and $ WW$ are, respectively, 
$\sim  4$ and $\sim 9$ times larger than that for the diphoton channel.  
Note that in many minimal models the signal is obtained by coupling a singlet scalar, which is unmixed with the SM Higgs, to vector fermions carrying only color and hypercharge.  Then, the $ZZ$ coupling to $\eta$ is loop induced and sub-dominant to the $\gamma \gamma$ coupling, 
due to suppression by $\tan^2 \theta_W$, where $\theta_W$ is the weak mixing angle.  In such models, one does not expect any significant branching ratio into the $WW$ final state.  

The presence of significant $ZZ$ and $WW$ branching fractions for $\eta$ in our model can then provide an interesting 
signal of Higgs-$\eta$ mixing that can be accessible in the LHC Run 2.  
In particular, a measurement of the ratio of 
BR$(\eta\to ZZ)$ to BR$(\eta\to \gamma \gamma)$, given the value of the diphoton signal strength, can yield 
the amount of Higgs-$\eta$ mixing. 
Also, if the coupling of $\eta$ to $Z$ and $W$ is dominated by the Higgs portal, then in general
\bea
\frac{\text{BR}(\eta\to W W)}{\text{BR}(\eta\to Z Z)}\approx 2.
\eea
This can then provide a test of our assumption regarding the induced Higgs mass parameter  
from $\vev{\eta} \neq 0$.

With our benchmark parameters, the mass of the two $\Qoppa$'s in the third generation,
$m_\Qoppa^>$, is set heavy to satisfy the bound from vector-like quark
searches.  The $\Qoppa$'s decay through three channels, $\Qoppa\to t W^-$,
$\Qoppa\to b Z$, and $\Qoppa\to b H$, with branching ratios of $\sim50\%$,
$\sim25\%$, and $\sim25\%$, respectively.  For small mixing between $\Qoppa$
and quark, the individual decay rates are roughly proportional to
$\sin^2\theta_L^{\Qoppa}$, where $\theta_L^{\Qoppa}$ is the mixing angle
between $\Qoppa_L$ and $b_L$, so the three branching ratios are roughly
constant.  Given these values, the bound on vector-like quark masses is 790 GeV
\cite{Khachatryan:2015gza}, therefore we set $m_\Qoppa^>$ to be 800 GeV.

The masses of the $\Qoppa$'s in the first two generations have to be heavier than half of the $\eta$ mass, to avoid the
$\eta\to \Qoppa\bar\Qoppa$ decay suppressing the signal.
Apart from that, these lighter exo-quarks are not subject to severe constraints as they mainly decay to $Wj,Zj$ and $Hj$.
One relevant search
channel is stop pair production, with each stop decaying into a charm quark and
a neutralino \cite{CMS:2014yma}.  The exo-quarks could decay through $\Qoppa\to
Zj\to j$+MET which has the same signal, but 
for our benchmark parameters the cross sections are $1\sim2$ orders of magnitude below the
uncertainty from the background.

The $\Lambda$ masses do not have severe bounds either \cite{Aad:2015dha}, except that they
should be again larger than $m_\eta/2$.  We set them at 380 GeV, nearly
one half of the exo-Higgs mass, mainly to enhance the signal and keep $v_\eta$
well above 1 TeV.  This is however not a strict requirement. 
For example, $m_\Lambda\sim 420$ GeV is still feasible with a somewhat larger $g_e$.

Once $v_\eta$ is fixed, $\lambda_\eta=0.20$ can be derived, and 
Eq.~(\ref{fopt-cond}) requires $g_e>1.48$.  We choose $g_e=2$ to 
be well inside the region of parameters favored by a strong first 
order phase transition.  This value then sets the mass of $\omega$ vector 
bosons, $m_\omega= 1.2$ TeV.  As
we will see later, with this mass $\omega$ production at LHC 7 and
8 TeV runs is too suppressed to yield a significant signal.  However, the Run 2 of the LHC
will have a chance to discover the $\omega$ bosons.

The mixing angles between $\Qoppa$'s and quarks should be well below
$\mathcal{O}(1\%)$, so that $\eta\to q\bar\Qoppa,\Qoppa\bar{q}$ decay rates
will not affect the signal strength too much.  
Apart from this condition,
the phenomenology does not depend
much on the value of these mixing angles. 
Since the $\Qoppa$ mixing sector
has two kinds of mixing terms, $\bar\Qoppa_Lq_R$ and $\bar q_L\Qoppa_R$,
the left- and right-handed mixing angles are in principle independent of each
other. We notice that for a mixing angle less than $O(10^{-7})$, the exo-quark would decay with a displaced vertex.
We choose $\theta^\Qoppa_{L,R}= 10^{-3}$ for simplicity.

The mixing between $\Lambda$'s and leptons is different: the $\mul$ terms could
couple $\Lambda_L$ to $l_R$, but mass couplings from $\Lambda_R$ to $l_L$ are
not allowed.  As a result, the left handed mixing angle, $\theta^\Lambda_L$, is
suppressed by the lepton mass:
\begin{equation}
	\frac{\tan\theta^\Lambda_L}{\tan\theta^\Lambda_R}=\frac{m_l}{m_\Lambda}.
\end{equation}
We choose $\mul=1$ GeV, corresponding to $\theta_R^\Lambda\approx2.7\times
10^{-3}$ for all three generations. Note that with this choice, the left-handed
mixing angle between the electron and $\Lambda^{\spup,\sdown;1}$ is extremely
tiny, $\sim3.6\times 10^{-9}$.   If $\mul\gsim10$ GeV the signal strength will
be affected at $\mathcal{O}(10\%)$ level by $\eta\to l\Lambda$, while if
$\mul\lsim0.1$ GeV $\Lambda$ will decay with a displaced vertex.

\begin{figure}[t]
\begin{center}
\includegraphics[width=0.48\textwidth]{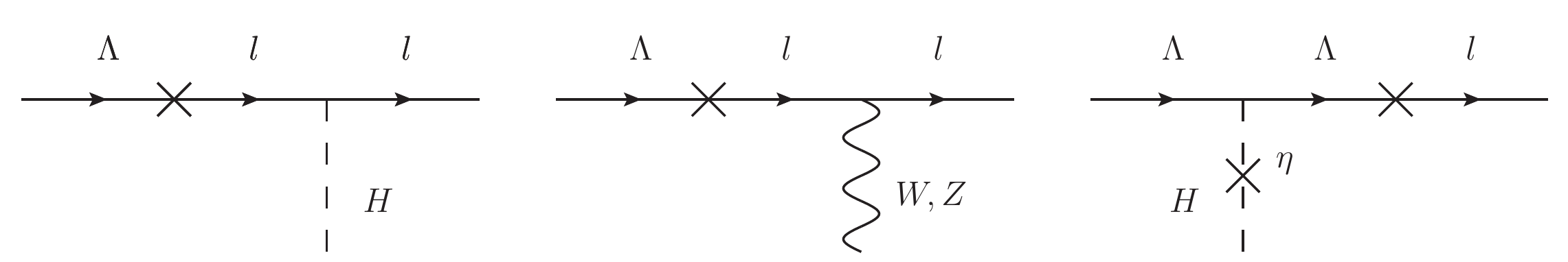}
\end{center}
\caption{Decays of a $\Lambda$ into a leptons. The decay into $\tau$ is mostly
	mediated by the first two processes, while the decays into $\mu$ and
	$e$ are mostly mediated by the third, since the first two are suppressed
	by $m_e$ and $m_\mu$.
}
\label{fig:LambdaDecay}
\end{figure}

With our choice of benchmark parameters, the dominant decay channel for $\Lambda^{\spup,\sdown;1}$
and $\Lambda^{\spup,\sdown;2}$ is $\Lambda^{\spup,\sdown;i}\to H+l_i$ where $i=1,2$ is the flavor
index.  If $\theta_{\eta H}=0$, the branching ratios of
the three possible decay channels, $\Lambda^{\spup,\sdown;i}\to W^-+\nu_i$,
$\Lambda^{\spup,\sdown;i}\to Z+l_i$, and $\Lambda^{\spup,\sdown;i}\to H+l_i$, are roughly $50\%$,
$25\%$, and $25\%$, respectively.
However the first two decay modes, $W^-+\nu$ and $Z+l$, occur through only
left-handed mixing $\theta^\Lambda_L$, which is suppressed by $m_l/m_\Lambda$.
This is because the right handed $\Lambda$ and $l$ only couple to the
hypercharge boson, with the same hypercharge, so a unitary rotation between
$\Lambda$ and $l$ will not generate any off-diagonal coupling.  On the other
hand, the last decay channel $\Lambda\to H+l$ occurs through the first diagram in
Fig.~\ref{fig:LambdaDecay}, and is also suppressed by $m_l$ because of the
lepton Yukawa.  As a result the decay rate of $\Lambda^{\spup,\sdown;1}$ is below
$10^{-13}$ GeV and would lead to displaced vertex.  However, with a nonzero
$\theta_{\eta H}$, the leptons can decay through the last diagram in
Fig.~\ref{fig:LambdaDecay}.  Even though the diagram involves two mixing angles
$\theta_R^\Lambda$ and $\theta_{\eta H}$, for electron and muon this is still
the dominant channel.  As a result, $\Lambda^{\spup,\sdown;1,2}$ mostly decay to Higgs and
lepton.  The branching ratios of all three exo-leptons are given in
Table~\ref{tab:lambdadecay}.  

\begin{table}
	\begin{tabular}{cccc}
		Mode & $\Lambda^{\spup,\sdown;1}$ & $\Lambda^{\spup,\sdown;2}$ & $\Lambda^{\spup,\sdown;3}$
		\\\hline
		$Hl$ & 100\%      &  84.0\%  &  22.0\%
		\\
		$W^-\nu$ & $\sim 10^{-6}$ &  10.6\% &  52.1\%
		\\
		$Zl$ & $\sim 10^{-6}$ &   5.3\%  & 25.9\%
		\\
	\end{tabular}
	\caption{$\Lambda$ branching ratios.		\label{tab:lambdadecay}}
\end{table}

It is interesting to see that the lighter lepton has a larger branching
ratio to the Higgs.  The corresponding rate for $\Lambda^{\spup,\sdown;1}$ is about
$7\times 10^{-11}$ GeV, above the limit for displaced vertex.

Our assumption, that $v_\eta$ leads to the electroweak symmetry breaking
of the SM, gives the right amount of $\eta-H$ mixing, that is enough to
give a reasonable decay rate for $\Lambda$, but is not too large to dilute the
signal through the Higgs portal.

\section{Predictions}

In this section, we will discuss the predictions and collider signals of our model 
that can be looked for in the future.  Within our mode, we predict 
new heavy quarks and leptons, i.e., $\Qoppa$ and $\Lambda$,
that are vector-like under the SM symmetries but chiral under the exo group.
The decay modes of $\Qoppa$'s are $\Qoppa^{\spup,\sdown;3}\to bZ,bH,tW^-$,
$\Qoppa^{\spup,\sdown;1,2}\to jZ,jH,jW^-$, and of 
$\Lambda$'s are $lZ,lH,\nu_l W^-$.  In this sense they are fairly standard
vector-like fermions, and can be discovered in corresponding searches. This is
similar to many other models that use vector-like fermions to explain the
750 GeV resonance.

The more distinct signature of our model is the production of the exo-gauge
bosons, i.e.~$\omega^\sraise$, $\omega^\slower$ (defined respectively as $(\omega^1- i\omega^2)/\sqrt{2}$ and $(\omega^1+ i\omega^2)/\sqrt{2}$) and
$\omega^3$.  The main production channel is
through $\Qoppa$-loop induced processes, $gg\to \omega\omega$ and $gg\to \omega
j$.  Note that the latter actually vanishes with our choice of parameters: the
$\Qoppa^\spup$ and $\Qoppa^\sdown$ masses are set to be equal, but the loop has a trace
over the $T^3$ generator in the exo-group, so their contributions cancel each
other.  Therefore to give a reasonable estimate on the cross section, we
increase the $\Qoppa^\spup$ masses by 300 GeV, only for this process.  The cross
sections we found are in Table~\ref{tab:omega}.  Note that the exo-gauge
bosons could be produced also at the tree level, through $q\bar q\to \omega$,
or $q g\to\omega\Qoppa$.  However the first is suppressed by
$({\theta_{L,R}^\Qoppa})^4$ and the second is by $({\theta_{L,R}^\Qoppa})^2$, and
the resulting cross sections are negligible with
${\theta_{L,R}^\Qoppa}\lsim 10^{-2}$.  
\begin{table}[h!]
	\begin{tabular}{cccc}
Process & 8 TeV & 13 TeV & 14 TeV
\\\hline
$gg\to\omega^3j$ & 0.016 fb & 0.16 fb & 0.26 fb
\\
$gg\to\omega\omega$ & 0.003 fb & 0.11 fb & 0.17 fb
	\end{tabular}
	\caption{Cross sections of the main production channels at different
		energies. The double omega production sums over all exo-vector bosons.\label{tab:omega}}
\end{table}

From Table~\ref{tab:omega}, we can see that currently available LHC data
should contain less than one $\omega$; at 14 TeV, however, with 300 fb$^{-1}$ accumulated luminosity
we can have $\mathcal{O}(10^2)$ $\omega$'s produced.
Even though the $\omega^3j$ production has a higher cross section, it depends
on the mass splitting.  We therefore focus on the pair production of $\omega^3$
and $\omega^\sraiselower$, which is our robust prediction.  Consider the leptonic
decay channel of $\omega$, {\it e.g.} $\omega^3\to\Lambda^{\spup,\sdown;i+}\Lambda^{\spup,\sdown;i-}$, where
$i=1,2$ is the flavor index, the total branching ratio is 28.8\%; $\omega^\sraiselower$
will decay into $\Lambda^{\spup,\sdown;+}\Lambda^{\sdown,\spup; -}$, but will not change the
counting. 
Using the cross section given in Table~\ref{tab:omega},
we will end up with $0.17\times300\times 28.8\%^2=4.2$ events with four
$\Lambda$'s given 300 fb$^{-1}$ of accumulated luminosity .  Among these events, $\Lambda^{\spup,\sdown;1}$ decays to the electron and
Higgs boson $\sim100\%$ of the time, while $\Lambda^{\spup,\sdown;2}$ decays to
the muon plus $H/Z$ with $\sim90\%$ branching ratio in total.  Therefore we will
have 3.5 signal events containing four leptons coming from the decay of
$\Lambda$'s, with a typical $p_T$ around $100\sim200$ GeV, and some additional
jets or leptons from $H$ or $Z$ decays.

These events with four hard leptons are significant enough that they 
will not be missed.  We estimate that the
irreducible SM background, from four leptons with four Higgs/$Z$ bosons, is
negligible.  Other background sources, for example those from 
$tt\bar t\bar t$ in all-leptonic channels, can be removed by requiring 
$p_T(l)>p_{Tcut}$, and $p_{Tcut}\approx 80$ GeV can already remove more than
$98\%$ of the background (that is with less than 0.1 event left), while reducing the
signal by about 30\%.  The analysis can be further elaborated by requiring no
missing transverse energy or requiring additional $j/l$, which will bring down the background by
another $1\sim2$ orders of magnitude.  In general, $4l$ production
from the SM, after removing opposite-sign same-flavor lepton pairs coming from $Z$'s and requiring
$p_T(l)>80$ GeV, is also below one event, and will become negligible once
additional jets are required.

We can also consider tagging the $b$'s from Higgs decay, which makes our
signal even more distinguishable. Considering $\Lambda\to lH$ only, we will have
about 3.1 $4H4l$ events at 300 fb$^{-1}$. If we require two of the four Higgs
bosons decay to $b\bar b$ and tag four $b$'s, we will have $2\sim3$ signal
events with 4 $b$'s, 4 hard leptons, and additional jets/leptons from Higgs
decay.  This is an even more distinct signal that can discriminate our model
from other new physics scenarios, and is essentially free of background.  

Of course, if the 750 GeV signal is
confirmed, dedicated analysis will be needed to optimize the search strategy
and to give a reliable estimate of the discovery potential, but the simple
estimation described above already shows that the $\omega$ pair production is a
promising channel.

\section{Discussions\label{sec:disc}}

\begin{figure}[t]
\begin{center}
\includegraphics[width=0.45\textwidth]{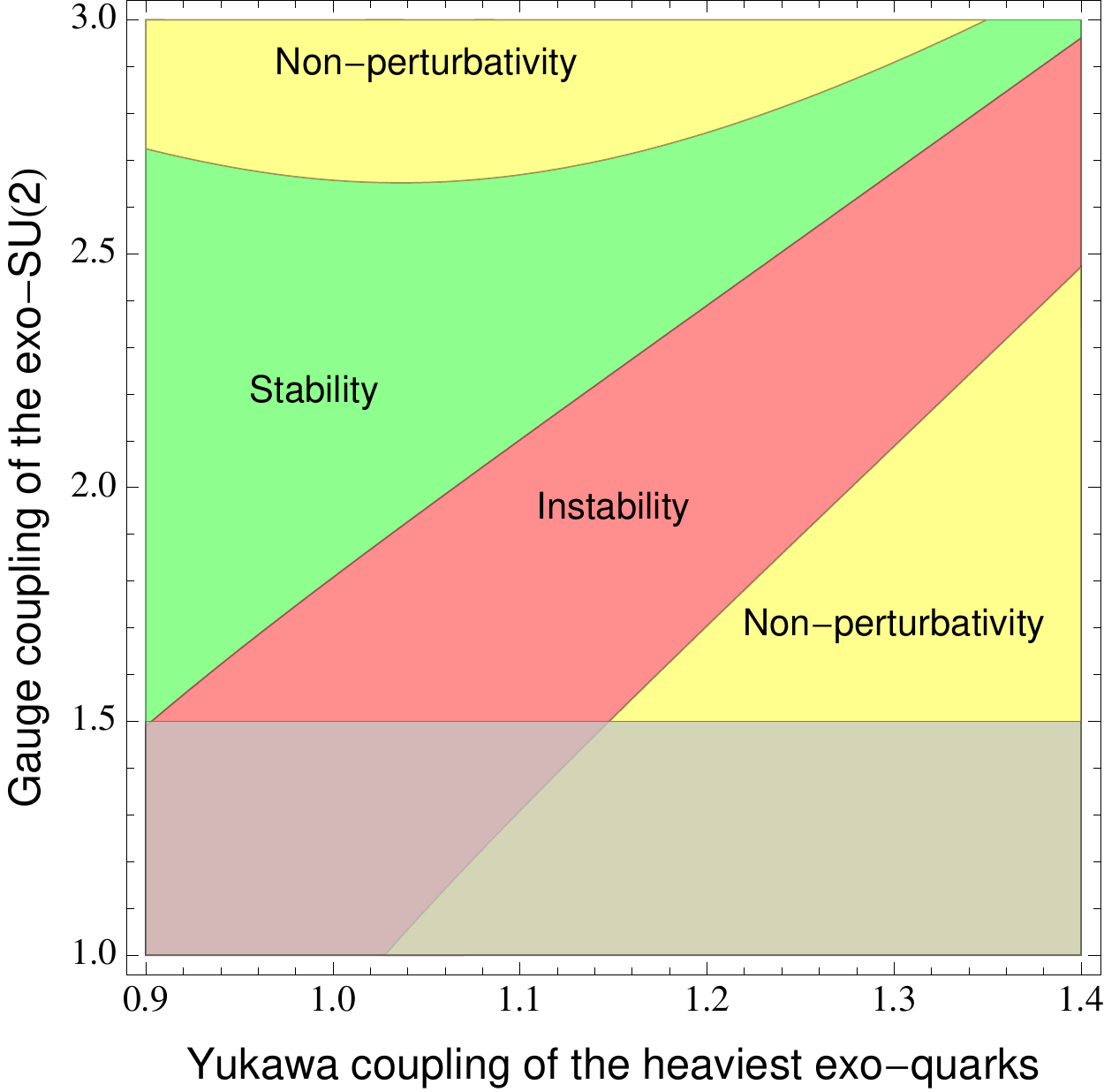}
\end{center}
\caption{The green area represents points in the $y_\Qoppa-g_e$ plane where all the parameters of the model stay positive and perturbative up to $\bar{\mu}=10^5\tev$. For the points in the red area $\lambda_\eta$ crosses zero before  $\bar{\mu}$. The points in the yellow region give a Landau-Pole for either $\lambda_\eta$ or $y_\Qoppa$. The grey area is excluded if we require a first-order transition}
\label{fig:stability}
\end{figure}

The model we have considered in this work contains a number of new fields that have significant interaction strengths.  
Hence, one may worry about quantum effects  of these fields on the validity of the underlying model. 
We will denote by $\bar\mu$ the maximum energy scale beyond which our model would need further completion to avoid loss of theoretical  control.  In Fig.~\ref{fig:stability},
we present various regimes of the model in the $y_\Qoppa-g_e$ plane, where 
$y_\Qoppa$ is the Yukawa coupling of the heaviest exo-quarks (800~GeV in our benchmark set of parameters).  Here, we choose  $\bar\mu = 10^5$~TeV, for which any unwanted effects of higher dimension operators from ultraviolet (UV) completions of our model are expected to be quite suppressed. 
   
For $\bar\mu=10^5$~TeV, our model maintains stability ($\lambda_\eta >0$) 
and perturbative reliability (no Landau poles) in the green shaded area (``Stability").  The red area (``Instability") has 
$\lambda_\eta<0$ and leads to an unstable exo-Higgs potential, whereas the yellow region (``Non-perturbativity") entails 
Landau poles for either $\lambda_\eta$ or $y_\Qoppa$.  The lower part of the plot, the horizontal band 
shaded gray, is disfavored if one requires 
a strong first order $SU(2)_e$ phase transition. 

The region of ``Stability" (green), as can be seen from Fig.~\ref{fig:stability}, represents a significant 
part of the parameter space and does not require very special choices for a reliable model.  It is also interesting that 
this region basically coincides with that favored by the requirement of a strong first order phase transition, as motivated 
by an explanation of the baryon asymmetry in the universe.  This is related to the fact that the stability of the exo-Higgs 
quartic coupling gets enhanced by the contributions of $\omega$ gauge fields, which grow with larger $g_e$.  We also 
add that gravitational waves corresponding to a strong phase transition at 
a temperature $T\sim 1$~TeV, as assumed in our scenario for baryogenesis,  
are typically predicted to be observable by future space-based gravitational waved detectors \cite{Grojean:2006bp}, such as 
LISA (for recent work on this topic, see for example Ref.~\cite{Schwaller:2015tja}).

\section{Conclusions}

In this work, we have proposed that the diphoton excess at 750~GeV, reported by the ATLAS and CMS collaborations, can be
due to a scalar resonance that is the remnant of an $SU(2)_e$  exo-spin gauge symmetry breaking through 
the vev of an exo-Higgs doublet.  We assume that there are exo-fermions,  carrying
SM color and hypercharge, that get their masses from the exo-Higgs mechanism and mediate the gluon fusion production
and diphoton decays of the scalar. 

We choose the matter content (exo-quarks and exo-leptons) and their associated
quantum numbers such that $B-L$ is anomalous under $SU(2)_e$.  Hence, 
with the assumption of a strong first order phase transition, one may expect the generation of a non-zero $B-L$ asymmetry 
in the early universe that can be the origin of the cosmic baryon asymmetry.  
This mechanism will then have the advantage of being testable at the
TeV energies available at the LHC and future colliders, 
unlike those scenarios that originate from much higher scales.  We have also assumed that the coupling 
of the exo-Higgs and SM Higgs doublets leads to the generation of the SM Higgs mass parameter, once $SU(2)_e$ 
is broken.  This can explain the similar sizes of the exo-Higgs and SM Higgs mass scales, and allows $\eta$ to have tree level decays into $tt$,  $WW$ and $ZZ$.

While the main ingredients of our model employed in explaining the diphoton excess effectively resemble those of models with
a singlet scalar and vector-like fermions, the presence of TeV-scale vector bosons, corresponding to the broken generators of
$SU(2)_e$, is a distinct prediction of our proposal.  We find that double vector boson production, the most robust
prediction of our scenario, can lead to a discovery of these states with about 300~fb$^{-1}$ at the 14 TeV LHC.  The decay
of each vector boson dominantly produces two hard leptons, as well as two or more
$b$-jets and more leptons and light jets, and can hence yield signals that are
effectively background free.  Under some mild assumptions about the spectrum of
the model, single exo-vector boson production is also a viable discovery
channel in the LHC Run 2.  

The model we have studied can be valid up to very large scales of $\sim 10^5$~TeV, for typical choices of parameters.  This scale 
is sufficiently large that any unwanted contributions from a UV completion can be negligibly suppressed.  We also pointed out that 
the requirement of strong phase transition at a temperature of order 
1 TeV implies gravitational wave signals for our model that may be 
potentially detectable by future space-based missions, such as LISA.       

{\it Speramus Naturam, vel lectorem, notionibus nostris benignam esse.}

\section*{Acknowledgements}

We would like to thank Frank Paige for valuable discussions on collider
signals.  This work is supported in part by the United States Department of
Energy under Grant Contracts DE-SC0012704.

\bibliography{bib}

\begin{thebibliography}{34}%
\makeatletter
\providecommand \@ifxundefined [1]{%
 \@ifx{#1\undefined}
}%
\providecommand \@ifnum [1]{%
 \ifnum #1\expandafter \@firstoftwo
 \else \expandafter \@secondoftwo
 \fi
}%
\providecommand \@ifx [1]{%
 \ifx #1\expandafter \@firstoftwo
 \else \expandafter \@secondoftwo
 \fi
}%
\providecommand \natexlab [1]{#1}%
\providecommand \enquote  [1]{``#1''}%
\providecommand \bibnamefont  [1]{#1}%
\providecommand \bibfnamefont [1]{#1}%
\providecommand \citenamefont [1]{#1}%
\providecommand \href@noop [0]{\@secondoftwo}%
\providecommand \href [0]{\begingroup \@sanitize@url \@href}%
\providecommand \@href[1]{\@@startlink{#1}\@@href}%
\providecommand \@@href[1]{\endgroup#1\@@endlink}%
\providecommand \@sanitize@url [0]{\catcode `\\12\catcode `\$12\catcode
  `\&12\catcode `\#12\catcode `\^12\catcode `\_12\catcode `\%12\relax}%
\providecommand \@@startlink[1]{}%
\providecommand \@@endlink[0]{}%
\providecommand \url  [0]{\begingroup\@sanitize@url \@url }%
\providecommand \@url [1]{\endgroup\@href {#1}{\urlprefix }}%
\providecommand \urlprefix  [0]{URL }%
\providecommand \Eprint [0]{\href }%
\providecommand \doibase [0]{http://dx.doi.org/}%
\providecommand \selectlanguage [0]{\@gobble}%
\providecommand \bibinfo  [0]{\@secondoftwo}%
\providecommand \bibfield  [0]{\@secondoftwo}%
\providecommand \translation [1]{[#1]}%
\providecommand \BibitemOpen [0]{}%
\providecommand \bibitemStop [0]{}%
\providecommand \bibitemNoStop [0]{.\EOS\space}%
\providecommand \EOS [0]{\spacefactor3000\relax}%
\providecommand \BibitemShut  [1]{\csname bibitem#1\endcsname}%
\let\auto@bib@innerbib\@empty
\bibitem [{\citenamefont {{The ATLAS Collaboration}}(2015)}]{ATLAS}%
  \BibitemOpen
  \bibfield  {author} {\bibinfo {author} {\bibnamefont {{The ATLAS
  Collaboration}}},\ }\bibfield  {title} {\emph {\bibinfo {title} {{Search for
  resonances decaying to photon pairs in 3.2 fb$^{-1}$ of $pp$ collisions at
  $\sqrt{s}$ = 13 TeV with the ATLAS detector}}},\ }\href@noop {} {\  (\bibinfo
  {year} {2015})}\BibitemShut {NoStop}%
\bibitem [{\citenamefont {{CMS Collaboration}}(2016)}]{CMS:2016owr}%
  \BibitemOpen
  \bibfield  {author} {\bibinfo {author} {\bibnamefont {{CMS Collaboration}}},\
  }\bibfield  {title} {\emph {\bibinfo {title} {{Search for new physics in high
  mass diphoton events in $3.3~\mathrm{fb}^{-1}$ of proton-proton collisions at
  $\sqrt{s}=13~\mathrm{TeV}$ and combined interpretation of searches at
  $8~\mathrm{TeV}$ and $13~\mathrm{TeV}$}}},\ }\href@noop {} {\  (\bibinfo
  {year} {2016})}\BibitemShut {NoStop}%
\bibitem [{\citenamefont {Knapen}\ \emph {et~al.}(2016)\citenamefont {Knapen},
  \citenamefont {Melia}, \citenamefont {Papucci},\ and\ \citenamefont
  {Zurek}}]{Knapen:2015dap}%
  \BibitemOpen
  \bibfield  {author} {\bibinfo {author} {\bibfnamefont {S.}~\bibnamefont
  {Knapen}}, \bibinfo {author} {\bibfnamefont {T.}~\bibnamefont {Melia}},
  \bibinfo {author} {\bibfnamefont {M.}~\bibnamefont {Papucci}}, \ and\
  \bibinfo {author} {\bibfnamefont {K.}~\bibnamefont {Zurek}},\ }\bibfield
  {title} {\emph {\bibinfo {title} {{Rays of light from the LHC}}},\ }\href
  {\doibase 10.1103/PhysRevD.93.075020} {\bibfield  {journal} {\bibinfo
  {journal} {Phys. Rev.}\ }\textbf {\bibinfo {volume} {D93}},\ \bibinfo {pages}
  {075020} (\bibinfo {year} {2016})},\ \Eprint
  {http://arxiv.org/abs/1512.04928}{arXiv:1512.04928 [hep-ph]}\BibitemShut
  {NoStop}%
\bibitem [{\citenamefont {Franceschini}\ \emph {et~al.}(2016)\citenamefont
  {Franceschini}, \citenamefont {Giudice}, \citenamefont {Kamenik},
  \citenamefont {McCullough}, \citenamefont {Pomarol}, \citenamefont
  {Rattazzi}, \citenamefont {Redi}, \citenamefont {Riva}, \citenamefont
  {Strumia},\ and\ \citenamefont {Torre}}]{Franceschini:2015kwy}%
  \BibitemOpen
  \bibfield  {author} {\bibinfo {author} {\bibfnamefont {R.}~\bibnamefont
  {Franceschini}}, \bibinfo {author} {\bibfnamefont {G.~F.}\ \bibnamefont
  {Giudice}}, \bibinfo {author} {\bibfnamefont {J.~F.}\ \bibnamefont
  {Kamenik}}, \bibinfo {author} {\bibfnamefont {M.}~\bibnamefont {McCullough}},
  \bibinfo {author} {\bibfnamefont {A.}~\bibnamefont {Pomarol}}, \bibinfo
  {author} {\bibfnamefont {R.}~\bibnamefont {Rattazzi}}, \bibinfo {author}
  {\bibfnamefont {M.}~\bibnamefont {Redi}}, \bibinfo {author} {\bibfnamefont
  {F.}~\bibnamefont {Riva}}, \bibinfo {author} {\bibfnamefont {A.}~\bibnamefont
  {Strumia}}, \ and\ \bibinfo {author} {\bibfnamefont {R.}~\bibnamefont
  {Torre}},\ }\bibfield  {title} {\emph {\bibinfo {title} {{What is the $\gamma
  \gamma$ resonance at 750 GeV?}}},\ }\href {\doibase 10.1007/JHEP03(2016)144}
  {\bibfield  {journal} {\bibinfo  {journal} {JHEP}\ }\textbf {\bibinfo
  {volume} {03}},\ \bibinfo {pages} {144} (\bibinfo {year} {2016})},\ \Eprint
  {http://arxiv.org/abs/1512.04933}{arXiv:1512.04933 [hep-ph]}\BibitemShut
  {NoStop}%
\bibitem [{\citenamefont {McDermott}\ \emph {et~al.}(2016)\citenamefont
  {McDermott}, \citenamefont {Meade},\ and\ \citenamefont
  {Ramani}}]{McDermott:2015sck}%
  \BibitemOpen
  \bibfield  {author} {\bibinfo {author} {\bibfnamefont {S.~D.}\ \bibnamefont
  {McDermott}}, \bibinfo {author} {\bibfnamefont {P.}~\bibnamefont {Meade}}, \
  and\ \bibinfo {author} {\bibfnamefont {H.}~\bibnamefont {Ramani}},\
  }\bibfield  {title} {\emph {\bibinfo {title} {{Singlet Scalar Resonances and
  the Diphoton Excess}}},\ }\href {\doibase 10.1016/j.physletb.2016.02.033}
  {\bibfield  {journal} {\bibinfo  {journal} {Phys. Lett.}\ }\textbf {\bibinfo
  {volume} {B755}},\ \bibinfo {pages} {353} (\bibinfo {year} {2016})},\ \Eprint
  {http://arxiv.org/abs/1512.05326}{arXiv:1512.05326 [hep-ph]}\BibitemShut
  {NoStop}%
\bibitem [{\citenamefont {Falkowski}\ \emph {et~al.}(2016)\citenamefont
  {Falkowski}, \citenamefont {Slone},\ and\ \citenamefont
  {Volansky}}]{Falkowski:2015swt}%
  \BibitemOpen
  \bibfield  {author} {\bibinfo {author} {\bibfnamefont {A.}~\bibnamefont
  {Falkowski}}, \bibinfo {author} {\bibfnamefont {O.}~\bibnamefont {Slone}}, \
  and\ \bibinfo {author} {\bibfnamefont {T.}~\bibnamefont {Volansky}},\
  }\bibfield  {title} {\emph {\bibinfo {title} {{Phenomenology of a 750 GeV
  Singlet}}},\ }\href {\doibase 10.1007/JHEP02(2016)152} {\bibfield  {journal}
  {\bibinfo  {journal} {JHEP}\ }\textbf {\bibinfo {volume} {02}},\ \bibinfo
  {pages} {152} (\bibinfo {year} {2016})},\ \Eprint
  {http://arxiv.org/abs/1512.05777}{arXiv:1512.05777 [hep-ph]}\BibitemShut
  {NoStop}%
\bibitem [{\citenamefont {Aloni}\ \emph {et~al.}(2015)\citenamefont {Aloni},
  \citenamefont {Blum}, \citenamefont {Dery}, \citenamefont {Efrati},\ and\
  \citenamefont {Nir}}]{Aloni:2015mxa}%
  \BibitemOpen
  \bibfield  {author} {\bibinfo {author} {\bibfnamefont {D.}~\bibnamefont
  {Aloni}}, \bibinfo {author} {\bibfnamefont {K.}~\bibnamefont {Blum}},
  \bibinfo {author} {\bibfnamefont {A.}~\bibnamefont {Dery}}, \bibinfo {author}
  {\bibfnamefont {A.}~\bibnamefont {Efrati}}, \ and\ \bibinfo {author}
  {\bibfnamefont {Y.}~\bibnamefont {Nir}},\ }\bibfield  {title} {\emph
  {\bibinfo {title} {{On a possible large width 750 GeV diphoton resonance at
  ATLAS and CMS}}},\ }\href@noop {} {\  (\bibinfo {year} {2015})},\ \Eprint
  {http://arxiv.org/abs/1512.05778}{arXiv:1512.05778 [hep-ph]}\BibitemShut
  {NoStop}%
\bibitem [{\citenamefont {Angelescu}\ \emph {et~al.}(2016)\citenamefont
  {Angelescu}, \citenamefont {Djouadi},\ and\ \citenamefont
  {Moreau}}]{Angelescu:2015uiz}%
  \BibitemOpen
  \bibfield  {author} {\bibinfo {author} {\bibfnamefont {A.}~\bibnamefont
  {Angelescu}}, \bibinfo {author} {\bibfnamefont {A.}~\bibnamefont {Djouadi}},
  \ and\ \bibinfo {author} {\bibfnamefont {G.}~\bibnamefont {Moreau}},\
  }\bibfield  {title} {\emph {\bibinfo {title} {{Scenarii for interpretations
  of the LHC diphoton excess: two Higgs doublets and vector-like quarks and
  leptons}}},\ }\href {\doibase 10.1016/j.physletb.2016.02.064} {\bibfield
  {journal} {\bibinfo  {journal} {Phys. Lett.}\ }\textbf {\bibinfo {volume}
  {B756}},\ \bibinfo {pages} {126} (\bibinfo {year} {2016})},\ \Eprint
  {http://arxiv.org/abs/1512.04921}{arXiv:1512.04921 [hep-ph]}\BibitemShut
  {NoStop}%
\bibitem [{\citenamefont {Gupta}\ \emph {et~al.}(2015)\citenamefont {Gupta},
  \citenamefont {J{\"a}ger}, \citenamefont {Kats}, \citenamefont {Perez},\ and\
  \citenamefont {Stamou}}]{Gupta:2015zzs}%
  \BibitemOpen
  \bibfield  {author} {\bibinfo {author} {\bibfnamefont {R.~S.}\ \bibnamefont
  {Gupta}}, \bibinfo {author} {\bibfnamefont {S.}~\bibnamefont {J{\"a}ger}},
  \bibinfo {author} {\bibfnamefont {Y.}~\bibnamefont {Kats}}, \bibinfo {author}
  {\bibfnamefont {G.}~\bibnamefont {Perez}}, \ and\ \bibinfo {author}
  {\bibfnamefont {E.}~\bibnamefont {Stamou}},\ }\bibfield  {title} {\emph
  {\bibinfo {title} {{Interpreting a 750 GeV Diphoton Resonance}}},\
  }\href@noop {} {\  (\bibinfo {year} {2015})},\ \Eprint
  {http://arxiv.org/abs/1512.05332}{arXiv:1512.05332 [hep-ph]}\BibitemShut
  {NoStop}%
\bibitem [{\citenamefont {Chang}(2016)}]{Chang:2015bzc}%
  \BibitemOpen
  \bibfield  {author} {\bibinfo {author} {\bibfnamefont {S.}~\bibnamefont
  {Chang}},\ }\bibfield  {title} {\emph {\bibinfo {title} {{A Simple $U(1)$
  Gauge Theory Explanation of the Diphoton Excess}}},\ }\href {\doibase
  10.1103/PhysRevD.93.055016} {\bibfield  {journal} {\bibinfo  {journal} {Phys.
  Rev.}\ }\textbf {\bibinfo {volume} {D93}},\ \bibinfo {pages} {055016}
  (\bibinfo {year} {2016})},\ \Eprint
  {http://arxiv.org/abs/1512.06426}{arXiv:1512.06426 [hep-ph]}\BibitemShut
  {NoStop}%
\bibitem [{\citenamefont {Kaneta}\ \emph {et~al.}(2015)\citenamefont {Kaneta},
  \citenamefont {Kang},\ and\ \citenamefont {Lee}}]{Kaneta:2015qpf}%
  \BibitemOpen
  \bibfield  {author} {\bibinfo {author} {\bibfnamefont {K.}~\bibnamefont
  {Kaneta}}, \bibinfo {author} {\bibfnamefont {S.}~\bibnamefont {Kang}}, \ and\
  \bibinfo {author} {\bibfnamefont {H.-S.}\ \bibnamefont {Lee}},\ }\bibfield
  {title} {\emph {\bibinfo {title} {{Diphoton excess at the LHC Run 2 and its
  implications for a new heavy gauge boson}}},\ }\href@noop {} {\  (\bibinfo
  {year} {2015})},\ \Eprint {http://arxiv.org/abs/1512.09129}{arXiv:1512.09129
  [hep-ph]}\BibitemShut {NoStop}%
\bibitem [{\citenamefont {Duerr}\ \emph {et~al.}(2016)\citenamefont {Duerr},
  \citenamefont {Fileviez~Pérez},\ and\ \citenamefont
  {Smirnov}}]{Duerr:2016eme}%
  \BibitemOpen
  \bibfield  {author} {\bibinfo {author} {\bibfnamefont {M.}~\bibnamefont
  {Duerr}}, \bibinfo {author} {\bibfnamefont {P.}~\bibnamefont
  {Fileviez~Pérez}}, \ and\ \bibinfo {author} {\bibfnamefont {J.}~\bibnamefont
  {Smirnov}},\ }\bibfield  {title} {\emph {\bibinfo {title} {{New Forces and
  the 750 GeV Resonance}}},\ }\href@noop {} {\  (\bibinfo {year} {2016})},\
  \Eprint {http://arxiv.org/abs/1604.05319}{arXiv:1604.05319
  [hep-ph]}\BibitemShut {NoStop}%
\bibitem [{\citenamefont {Harvey}\ and\ \citenamefont
  {Turner}(1990)}]{Harvey:1990qw}%
  \BibitemOpen
  \bibfield  {author} {\bibinfo {author} {\bibfnamefont {J.~A.}\ \bibnamefont
  {Harvey}}\ and\ \bibinfo {author} {\bibfnamefont {M.~S.}\ \bibnamefont
  {Turner}},\ }\bibfield  {title} {\emph {\bibinfo {title} {{Cosmological
  baryon and lepton number in the presence of electroweak fermion number
  violation}}},\ }\href {\doibase 10.1103/PhysRevD.42.3344} {\bibfield
  {journal} {\bibinfo  {journal} {Phys. Rev.}\ }\textbf {\bibinfo {volume}
  {D42}},\ \bibinfo {pages} {3344} (\bibinfo {year} {1990})}\BibitemShut
  {NoStop}%
\bibitem [{\citenamefont {Shu}\ \emph {et~al.}(2007)\citenamefont {Shu},
  \citenamefont {Tait},\ and\ \citenamefont {Wagner}}]{Shu:2006mm}%
  \BibitemOpen
  \bibfield  {author} {\bibinfo {author} {\bibfnamefont {J.}~\bibnamefont
  {Shu}}, \bibinfo {author} {\bibfnamefont {T.~M.~P.}\ \bibnamefont {Tait}}, \
  and\ \bibinfo {author} {\bibfnamefont {C.~E.~M.}\ \bibnamefont {Wagner}},\
  }\bibfield  {title} {\emph {\bibinfo {title} {{Baryogenesis from an Earlier
  Phase Transition}}},\ }\href {\doibase 10.1103/PhysRevD.75.063510} {\bibfield
   {journal} {\bibinfo  {journal} {Phys. Rev.}\ }\textbf {\bibinfo {volume}
  {D75}},\ \bibinfo {pages} {063510} (\bibinfo {year} {2007})},\ \Eprint
  {http://arxiv.org/abs/hep-ph/0610375}{arXiv:hep-ph/0610375}\BibitemShut
  {NoStop}%
\bibitem [{\citenamefont {Agashe}\ \emph {et~al.}(2010)\citenamefont {Agashe},
  \citenamefont {Kim}, \citenamefont {Toharia},\ and\ \citenamefont
  {Walker}}]{Agashe:2010gt}%
  \BibitemOpen
  \bibfield  {author} {\bibinfo {author} {\bibfnamefont {K.}~\bibnamefont
  {Agashe}}, \bibinfo {author} {\bibfnamefont {D.}~\bibnamefont {Kim}},
  \bibinfo {author} {\bibfnamefont {M.}~\bibnamefont {Toharia}}, \ and\
  \bibinfo {author} {\bibfnamefont {D.~G.~E.}\ \bibnamefont {Walker}},\
  }\bibfield  {title} {\emph {\bibinfo {title} {{Distinguishing Dark Matter
  Stabilization Symmetries Using Multiple Kinematic Edges and Cusps}}},\ }\href
  {\doibase 10.1103/PhysRevD.82.015007} {\bibfield  {journal} {\bibinfo
  {journal} {Phys. Rev.}\ }\textbf {\bibinfo {volume} {D82}},\ \bibinfo {pages}
  {015007} (\bibinfo {year} {2010})},\ \Eprint
  {http://arxiv.org/abs/1003.0899}{arXiv:1003.0899 [hep-ph]}\BibitemShut
  {NoStop}%
\bibitem [{\citenamefont {Walker}(2012)}]{Walker:2012ka}%
  \BibitemOpen
  \bibfield  {author} {\bibinfo {author} {\bibfnamefont {D.~G.~E.}\
  \bibnamefont {Walker}},\ }\bibfield  {title} {\emph {\bibinfo {title} {{Dark
  Baryogenesis}}},\ }\href@noop {} {\  (\bibinfo {year} {2012})},\ \Eprint
  {http://arxiv.org/abs/1202.2348}{arXiv:1202.2348 [hep-ph]}\BibitemShut
  {NoStop}%
\bibitem [{\citenamefont {Witten}(1982)}]{Witten:1982fp}%
  \BibitemOpen
  \bibfield  {author} {\bibinfo {author} {\bibfnamefont {E.}~\bibnamefont
  {Witten}},\ }\bibfield  {title} {\emph {\bibinfo {title} {{An SU(2)
  Anomaly}}},\ }\href {\doibase 10.1016/0370-2693(82)90728-6} {\bibfield
  {journal} {\bibinfo  {journal} {Phys. Lett.}\ }\textbf {\bibinfo {volume}
  {B117}},\ \bibinfo {pages} {324} (\bibinfo {year} {1982})}\BibitemShut
  {NoStop}%
\bibitem [{\citenamefont {Sakharov}(1967)}]{Sakharov:1967dj}%
  \BibitemOpen
  \bibfield  {author} {\bibinfo {author} {\bibfnamefont {A.~D.}\ \bibnamefont
  {Sakharov}},\ }\bibfield  {title} {\emph {\bibinfo {title} {{Violation of CP
  Invariance, c Asymmetry, and Baryon Asymmetry of the Universe}}},\ }\href
  {\doibase 10.1070/PU1991v034n05ABEH002497} {\bibfield  {journal} {\bibinfo
  {journal} {Pisma Zh. Eksp. Teor. Fiz.}\ }\textbf {\bibinfo {volume} {5}},\
  \bibinfo {pages} {32} (\bibinfo {year} {1967})},\ \bibinfo {note} {[Usp. Fiz.
  Nauk161,61(1991)]}\BibitemShut {NoStop}%
\bibitem [{\citenamefont {Fukugita}\ and\ \citenamefont
  {Yanagida}(1986)}]{Fukugita:1986hr}%
  \BibitemOpen
  \bibfield  {author} {\bibinfo {author} {\bibfnamefont {M.}~\bibnamefont
  {Fukugita}}\ and\ \bibinfo {author} {\bibfnamefont {T.}~\bibnamefont
  {Yanagida}},\ }\bibfield  {title} {\emph {\bibinfo {title} {{Baryogenesis
  Without Grand Unification}}},\ }\href {\doibase 10.1016/0370-2693(86)91126-3}
  {\bibfield  {journal} {\bibinfo  {journal} {Phys. Lett.}\ }\textbf {\bibinfo
  {volume} {B174}},\ \bibinfo {pages} {45} (\bibinfo {year}
  {1986})}\BibitemShut {NoStop}%
\bibitem [{\citenamefont {Quiros}(1999)}]{Quiros:1999jp}%
  \BibitemOpen
  \bibfield  {author} {\bibinfo {author} {\bibfnamefont {M.}~\bibnamefont
  {Quiros}},\ }\bibfield  {title} {\emph {\bibinfo {title} {{Finite temperature
  field theory and phase transitions}}},\ }in\ \href
  {http://alice.cern.ch/format/showfull?sysnb=0302087} {\emph {\bibinfo
  {booktitle} {{High energy physics and cosmology. Proceedings, Summer School,
  Trieste, Italy, June 29-July 17, 1998}}}}\ (\bibinfo {year} {1999})\ pp.\
  \bibinfo {pages} {187--259},\ \Eprint
  {http://arxiv.org/abs/hep-ph/9901312}{arXiv:hep-ph/9901312}\BibitemShut
  {NoStop}%
\bibitem [{\citenamefont {Alwall}\ \emph {et~al.}(2014)\citenamefont {Alwall},
  \citenamefont {Frederix}, \citenamefont {Frixione}, \citenamefont {Hirschi},
  \citenamefont {Maltoni} \emph {et~al.}}]{Alwall:2014hca}%
  \BibitemOpen
  \bibfield  {author} {\bibinfo {author} {\bibfnamefont {J.}~\bibnamefont
  {Alwall}}, \bibinfo {author} {\bibfnamefont {R.}~\bibnamefont {Frederix}},
  \bibinfo {author} {\bibfnamefont {S.}~\bibnamefont {Frixione}}, \bibinfo
  {author} {\bibfnamefont {V.}~\bibnamefont {Hirschi}}, \bibinfo {author}
  {\bibfnamefont {F.}~\bibnamefont {Maltoni}},  \emph {et~al.},\ }\bibfield
  {title} {\emph {\bibinfo {title} {{The automated computation of tree-level
  and next-to-leading order differential cross sections, and their matching to
  parton shower simulations}}},\ }\href {\doibase 10.1007/JHEP07(2014)079}
  {\bibfield  {journal} {\bibinfo  {journal} {JHEP}\ }\textbf {\bibinfo
  {volume} {1407}},\ \bibinfo {pages} {079} (\bibinfo {year} {2014})},\ \Eprint
  {http://arxiv.org/abs/1405.0301}{arXiv:1405.0301 [hep-ph]}\BibitemShut
  {NoStop}%
\bibitem [{\citenamefont {Alwall}\ \emph {et~al.}(2015)\citenamefont {Alwall},
  \citenamefont {Duhr}, \citenamefont {Fuks}, \citenamefont {Mattelaer},
  \citenamefont {Özturk},\ and\ \citenamefont {Shen}}]{Alwall:2014bza}%
  \BibitemOpen
  \bibfield  {author} {\bibinfo {author} {\bibfnamefont {J.}~\bibnamefont
  {Alwall}}, \bibinfo {author} {\bibfnamefont {C.}~\bibnamefont {Duhr}},
  \bibinfo {author} {\bibfnamefont {B.}~\bibnamefont {Fuks}}, \bibinfo {author}
  {\bibfnamefont {O.}~\bibnamefont {Mattelaer}}, \bibinfo {author}
  {\bibfnamefont {D.~G.}\ \bibnamefont {Özturk}}, \ and\ \bibinfo {author}
  {\bibfnamefont {C.-H.}\ \bibnamefont {Shen}},\ }\bibfield  {title} {\emph
  {\bibinfo {title} {{Computing decay rates for new physics theories with
  FeynRules and MadGraph5\_aMC@NLO}}},\ }\href {\doibase
  10.1016/j.cpc.2015.08.031} {\bibfield  {journal} {\bibinfo  {journal}
  {Comput. Phys. Commun.}\ }\textbf {\bibinfo {volume} {197}},\ \bibinfo
  {pages} {312} (\bibinfo {year} {2015})},\ \Eprint
  {http://arxiv.org/abs/1402.1178}{arXiv:1402.1178 [hep-ph]}\BibitemShut
  {NoStop}%
\bibitem [{\citenamefont {Degrande}\ \emph {et~al.}(2012)\citenamefont
  {Degrande}, \citenamefont {Duhr}, \citenamefont {Fuks}, \citenamefont
  {Grellscheid}, \citenamefont {Mattelaer},\ and\ \citenamefont
  {Reiter}}]{Degrande:2011ua}%
  \BibitemOpen
  \bibfield  {author} {\bibinfo {author} {\bibfnamefont {C.}~\bibnamefont
  {Degrande}}, \bibinfo {author} {\bibfnamefont {C.}~\bibnamefont {Duhr}},
  \bibinfo {author} {\bibfnamefont {B.}~\bibnamefont {Fuks}}, \bibinfo {author}
  {\bibfnamefont {D.}~\bibnamefont {Grellscheid}}, \bibinfo {author}
  {\bibfnamefont {O.}~\bibnamefont {Mattelaer}}, \ and\ \bibinfo {author}
  {\bibfnamefont {T.}~\bibnamefont {Reiter}},\ }\bibfield  {title} {\emph
  {\bibinfo {title} {{UFO - The Universal FeynRules Output}}},\ }\href
  {\doibase 10.1016/j.cpc.2012.01.022} {\bibfield  {journal} {\bibinfo
  {journal} {Comput. Phys. Commun.}\ }\textbf {\bibinfo {volume} {183}},\
  \bibinfo {pages} {1201} (\bibinfo {year} {2012})},\ \Eprint
  {http://arxiv.org/abs/1108.2040}{arXiv:1108.2040 [hep-ph]}\BibitemShut
  {NoStop}%
\bibitem [{\citenamefont {Alloul}\ \emph {et~al.}(2014)\citenamefont {Alloul},
  \citenamefont {Christensen}, \citenamefont {Degrande}, \citenamefont {Duhr},\
  and\ \citenamefont {Fuks}}]{Alloul:2013bka}%
  \BibitemOpen
  \bibfield  {author} {\bibinfo {author} {\bibfnamefont {A.}~\bibnamefont
  {Alloul}}, \bibinfo {author} {\bibfnamefont {N.~D.}\ \bibnamefont
  {Christensen}}, \bibinfo {author} {\bibfnamefont {C.}~\bibnamefont
  {Degrande}}, \bibinfo {author} {\bibfnamefont {C.}~\bibnamefont {Duhr}}, \
  and\ \bibinfo {author} {\bibfnamefont {B.}~\bibnamefont {Fuks}},\ }\bibfield
  {title} {\emph {\bibinfo {title} {{FeynRules 2.0 - A complete toolbox for
  tree-level phenomenology}}},\ }\href {\doibase 10.1016/j.cpc.2014.04.012}
  {\bibfield  {journal} {\bibinfo  {journal} {Comput. Phys. Commun.}\ }\textbf
  {\bibinfo {volume} {185}},\ \bibinfo {pages} {2250} (\bibinfo {year}
  {2014})},\ \Eprint {http://arxiv.org/abs/1310.1921}{arXiv:1310.1921
  [hep-ph]}\BibitemShut {NoStop}%
\bibitem [{\citenamefont {Hirschi}\ and\ \citenamefont
  {Mattelaer}(2015)}]{Hirschi:2015iia}%
  \BibitemOpen
  \bibfield  {author} {\bibinfo {author} {\bibfnamefont {V.}~\bibnamefont
  {Hirschi}}\ and\ \bibinfo {author} {\bibfnamefont {O.}~\bibnamefont
  {Mattelaer}},\ }\bibfield  {title} {\emph {\bibinfo {title} {{Automated event
  generation for loop-induced processes}}},\ }\href {\doibase
  10.1007/JHEP10(2015)146} {\bibfield  {journal} {\bibinfo  {journal} {JHEP}\
  }\textbf {\bibinfo {volume} {10}},\ \bibinfo {pages} {146} (\bibinfo {year}
  {2015})},\ \Eprint {http://arxiv.org/abs/1507.00020}{arXiv:1507.00020
  [hep-ph]}\BibitemShut {NoStop}%
\bibitem [{\citenamefont {Degrande}(2015)}]{Degrande:2014vpa}%
  \BibitemOpen
  \bibfield  {author} {\bibinfo {author} {\bibfnamefont {C.}~\bibnamefont
  {Degrande}},\ }\bibfield  {title} {\emph {\bibinfo {title} {{Automatic
  evaluation of UV and R2 terms for beyond the Standard Model Lagrangians: a
  proof-of-principle}}},\ }\href {\doibase 10.1016/j.cpc.2015.08.015}
  {\bibfield  {journal} {\bibinfo  {journal} {Comput. Phys. Commun.}\ }\textbf
  {\bibinfo {volume} {197}},\ \bibinfo {pages} {239} (\bibinfo {year}
  {2015})},\ \Eprint {http://arxiv.org/abs/1406.3030}{arXiv:1406.3030
  [hep-ph]}\BibitemShut {NoStop}%
\bibitem [{\citenamefont {Ball}\ \emph {et~al.}(2013)\citenamefont {Ball},
  \citenamefont {Bertone}, \citenamefont {Carrazza}, \citenamefont
  {Del~Debbio}, \citenamefont {Forte}, \citenamefont {Guffanti}, \citenamefont
  {Hartland},\ and\ \citenamefont {Rojo}}]{Ball:2013hta}%
  \BibitemOpen
  \bibfield  {author} {\bibinfo {author} {\bibfnamefont {R.~D.}\ \bibnamefont
  {Ball}}, \bibinfo {author} {\bibfnamefont {V.}~\bibnamefont {Bertone}},
  \bibinfo {author} {\bibfnamefont {S.}~\bibnamefont {Carrazza}}, \bibinfo
  {author} {\bibfnamefont {L.}~\bibnamefont {Del~Debbio}}, \bibinfo {author}
  {\bibfnamefont {S.}~\bibnamefont {Forte}}, \bibinfo {author} {\bibfnamefont
  {A.}~\bibnamefont {Guffanti}}, \bibinfo {author} {\bibfnamefont {N.~P.}\
  \bibnamefont {Hartland}}, \ and\ \bibinfo {author} {\bibfnamefont
  {J.}~\bibnamefont {Rojo}} (\bibinfo {collaboration} {NNPDF}),\ }\bibfield
  {title} {\emph {\bibinfo {title} {{Parton distributions with QED
  corrections}}},\ }\href {\doibase 10.1016/j.nuclphysb.2013.10.010} {\bibfield
   {journal} {\bibinfo  {journal} {Nucl. Phys.}\ }\textbf {\bibinfo {volume}
  {B877}},\ \bibinfo {pages} {290} (\bibinfo {year} {2013})},\ \Eprint
  {http://arxiv.org/abs/1308.0598}{arXiv:1308.0598 [hep-ph]}\BibitemShut
  {NoStop}%
\bibitem [{\citenamefont {Dulat}\ \emph {et~al.}(2016)\citenamefont {Dulat},
  \citenamefont {Hou}, \citenamefont {Gao}, \citenamefont {Guzzi},
  \citenamefont {Huston}, \citenamefont {Nadolsky}, \citenamefont {Pumplin},
  \citenamefont {Schmidt}, \citenamefont {Stump},\ and\ \citenamefont
  {Yuan}}]{Dulat:2015mca}%
  \BibitemOpen
  \bibfield  {author} {\bibinfo {author} {\bibfnamefont {S.}~\bibnamefont
  {Dulat}}, \bibinfo {author} {\bibfnamefont {T.-J.}\ \bibnamefont {Hou}},
  \bibinfo {author} {\bibfnamefont {J.}~\bibnamefont {Gao}}, \bibinfo {author}
  {\bibfnamefont {M.}~\bibnamefont {Guzzi}}, \bibinfo {author} {\bibfnamefont
  {J.}~\bibnamefont {Huston}}, \bibinfo {author} {\bibfnamefont
  {P.}~\bibnamefont {Nadolsky}}, \bibinfo {author} {\bibfnamefont
  {J.}~\bibnamefont {Pumplin}}, \bibinfo {author} {\bibfnamefont
  {C.}~\bibnamefont {Schmidt}}, \bibinfo {author} {\bibfnamefont
  {D.}~\bibnamefont {Stump}}, \ and\ \bibinfo {author} {\bibfnamefont {C.~P.}\
  \bibnamefont {Yuan}},\ }\bibfield  {title} {\emph {\bibinfo {title} {{New
  parton distribution functions from a global analysis of quantum
  chromodynamics}}},\ }\href {\doibase 10.1103/PhysRevD.93.033006} {\bibfield
  {journal} {\bibinfo  {journal} {Phys. Rev.}\ }\textbf {\bibinfo {volume}
  {D93}},\ \bibinfo {pages} {033006} (\bibinfo {year} {2016})},\ \Eprint
  {http://arxiv.org/abs/1506.07443}{arXiv:1506.07443 [hep-ph]}\BibitemShut
  {NoStop}%
\bibitem [{\citenamefont {Liu}\ and\ \citenamefont
  {Zhang}(2016)}]{Liu:2016mpd}%
  \BibitemOpen
  \bibfield  {author} {\bibinfo {author} {\bibfnamefont {X.}~\bibnamefont
  {Liu}}\ and\ \bibinfo {author} {\bibfnamefont {H.}~\bibnamefont {Zhang}},\
  }\bibfield  {title} {\emph {\bibinfo {title} {{RG-improved Prediction for 750
  GeV Resonance Production at the LHC}}},\ }\href@noop {} {\  (\bibinfo {year}
  {2016})},\ \Eprint {http://arxiv.org/abs/1603.07190}{arXiv:1603.07190
  [hep-ph]}\BibitemShut {NoStop}%
\bibitem [{\citenamefont {Khachatryan}\ \emph {et~al.}(2016)\citenamefont
  {Khachatryan} \emph {et~al.}}]{Khachatryan:2015gza}%
  \BibitemOpen
  \bibfield  {author} {\bibinfo {author} {\bibfnamefont {V.}~\bibnamefont
  {Khachatryan}} \emph {et~al.} (\bibinfo {collaboration} {CMS}),\ }\bibfield
  {title} {\emph {\bibinfo {title} {{Search for pair-produced vectorlike B
  quarks in proton-proton collisions at $\sqrt{s}$=8??TeV}}},\ }\href {\doibase
  10.1103/PhysRevD.93.112009} {\bibfield  {journal} {\bibinfo  {journal} {Phys.
  Rev.}\ }\textbf {\bibinfo {volume} {D93}},\ \bibinfo {pages} {112009}
  (\bibinfo {year} {2016})},\ \Eprint
  {http://arxiv.org/abs/1507.07129}{arXiv:1507.07129 [hep-ex]}\BibitemShut
  {NoStop}%
\bibitem [{\citenamefont {{CMS Collaboration}}(2014)}]{CMS:2014yma}%
  \BibitemOpen
  \bibfield  {author} {\bibinfo {author} {\bibnamefont {{CMS Collaboration}}},\
  }\bibfield  {title} {\emph {\bibinfo {title} {{Search for top squarks
  decaying to a charm quark and a neutralino in events with a jet and missing
  transverse momentum}}},\ }\href@noop {} {\  (\bibinfo {year}
  {2014})}\BibitemShut {NoStop}%
\bibitem [{\citenamefont {Aad}\ \emph {et~al.}(2015)\citenamefont {Aad} \emph
  {et~al.}}]{Aad:2015dha}%
  \BibitemOpen
  \bibfield  {author} {\bibinfo {author} {\bibfnamefont {G.}~\bibnamefont
  {Aad}} \emph {et~al.} (\bibinfo {collaboration} {ATLAS}),\ }\bibfield
  {title} {\emph {\bibinfo {title} {{Search for heavy lepton resonances
  decaying to a $Z$ boson and a lepton in $pp$ collisions at $\sqrt{s}=8$ TeV
  with the ATLAS detector}}},\ }\href {\doibase 10.1007/JHEP09(2015)108}
  {\bibfield  {journal} {\bibinfo  {journal} {JHEP}\ }\textbf {\bibinfo
  {volume} {09}},\ \bibinfo {pages} {108} (\bibinfo {year} {2015})},\ \Eprint
  {http://arxiv.org/abs/1506.01291}{arXiv:1506.01291 [hep-ex]}\BibitemShut
  {NoStop}%
\bibitem [{\citenamefont {Grojean}\ and\ \citenamefont
  {Servant}(2007)}]{Grojean:2006bp}%
  \BibitemOpen
  \bibfield  {author} {\bibinfo {author} {\bibfnamefont {C.}~\bibnamefont
  {Grojean}}\ and\ \bibinfo {author} {\bibfnamefont {G.}~\bibnamefont
  {Servant}},\ }\bibfield  {title} {\emph {\bibinfo {title} {{Gravitational
  Waves from Phase Transitions at the Electroweak Scale and Beyond}}},\ }\href
  {\doibase 10.1103/PhysRevD.75.043507} {\bibfield  {journal} {\bibinfo
  {journal} {Phys. Rev.}\ }\textbf {\bibinfo {volume} {D75}},\ \bibinfo {pages}
  {043507} (\bibinfo {year} {2007})},\ \Eprint
  {http://arxiv.org/abs/hep-ph/0607107}{arXiv:hep-ph/0607107}\BibitemShut
  {NoStop}%
\bibitem [{\citenamefont {Schwaller}(2015)}]{Schwaller:2015tja}%
  \BibitemOpen
  \bibfield  {author} {\bibinfo {author} {\bibfnamefont {P.}~\bibnamefont
  {Schwaller}},\ }\bibfield  {title} {\emph {\bibinfo {title} {{Gravitational
  Waves from a Dark Phase Transition}}},\ }\href {\doibase
  10.1103/PhysRevLett.115.181101} {\bibfield  {journal} {\bibinfo  {journal}
  {Phys. Rev. Lett.}\ }\textbf {\bibinfo {volume} {115}},\ \bibinfo {pages}
  {181101} (\bibinfo {year} {2015})},\ \Eprint
  {http://arxiv.org/abs/1504.07263}{arXiv:1504.07263 [hep-ph]}\BibitemShut
  {NoStop}%
\end{thebibliography}%
\bibliographystyle{apsrev4-1_title}

\end{document}